\documentclass[epjST]{svjour}
\usepackage{graphics}
\begin{document}
\title{Controlled generation of momentum states in a high-finesse
ring cavity}
\author{Nicola Piovella\inst{1}\fnmsep\thanks{\email{nicola.piovella@unimi.it}}}
\institute{Dipartimento di Fisica, Universit\`a degli Studi di
Milano, Via Celoria 16, Milano I-20133, Italy}
\abstract{ A Bose-Einstein condensate in a high-finesse ring
cavity scatters the photons of a pump beam into counterpropagating
cavity modes, populating a bi-dimensional momentum lattice. A
high-finesse ring cavity with a sub-recoil linewidth allows to
control the quantized atomic motion, selecting particular discrete
momentum states and generating atom-photon entanglement. The
semiclassical and quantum model for the 2D collective atomic
recoil lasing (CARL) are derived and the superradiant and
good-cavity regimes discussed. For pump incidence perpendicular to
the cavity axis, the momentum lattice is symmetrically populated.
Conversely, for oblique pump incidence the motion along the two
recoil directions is unbalanced and different momentum states can
be populated on demand by tuning the pump frequency.}

\maketitle

\section{Introduction}

Their unique coherence properties candidate  Bose-Einstein
condensates (BECs) as ideal systems to generate and probe
light-atom correlations in collective light scattering and
superradiant instabilities \cite{Yukalov}. Superradiant Rayleigh
scattering experiments usually occur in free space
\cite{Inouye1999,Fallani2005}, so that scattered photons rapidly
leave the interaction volume, limiting the coherence time of modes
propagating along the major condensate's axis ('end-fire modes').
On the contrary, when BECs interact with a high-finesse optical
cavity, the correlations between scattered events can be stored in
long-lived cavity modes \cite{SlamaPRL2007}, allowing, for
instance, to study new regimes in the strong coupling limit
\cite{Brennecke2007,Baumann2010}. Furthermore, recent experiments
on collective light scattering by BECs in a high-finesse ring
cavity have shown the possibility to employ the cavity sub-recoil
resolution as a filter selecting particular quantized momentum
states \cite{BuxPRL2011}. These experiments rely on the collective
atomic recoil lasing (CARL) mechanism, envisaged by Bonifacio and
coworker in 1994 \cite{CARL} and finally observed  in
T\"{u}bingen, early with atomic clouds as hot as several $100\mu$K
\cite{Kruse2003} and more recently with ultra-cold atoms
\cite{SlamaPRA2007}. CARL represents the atomic analogue of the
free-electron laser (FEL) \cite{Madey}, which has been studied for
a long time in Milan \cite{BPN,NC}. In particular, Bonifacio and
Casagrande predicted the existence of a superradiant regime in
FELs \cite{BC1984,BCJOSA1985}, successively observed as collective
light scattering in CARL \cite{SlamaPRA2007}. A further advance on
CARL theory was obtained in 2001, when the semiclassical model was
extended to a quantum description suitable for BECs, in which the
atomic motion is quantized in photon recoil momentum states
\cite{Gatelli2001,Martinucci2002}. Next, a full quantum CARL
theory investigated entanglement between collective momentum
states and cavity modes \cite{Piovella2003,Cola2004}. Generally,
CARL is often described in a 1D geometry, i.e. with pump and
scattered modes anti-parallel and atoms recoiling after each
scattering event by $2\hbar k$ along the incident beam direction.
However, in the former Superradiant Rayleigh scattering experiment
\cite{Inouye1999} the cigar-shaped condensate major axis was set
orthogonal to the incident laser, and two scattered beams were
emitted along the condensate axis, with atoms recoiling at
45$^\circ$ with respect to the incident laser. In that case, the
geometry was two-dimensional with two scattered end-fire modes.
CARL and Superradiant Rayleigh scattering in a 2D configuration
have been investigated by several authors
\cite{Moore1999,Muste2000,Trifonov2001,PiovellaLP2002,PiovellaLP2003,Zobay2005,Zobay2006,Hilliard2008,Lu2011}.
More recently, a two-frequency pumping scheme has been implemented
to enhance the resonant sequential scattering in the Superradiant
Rayleigh scattering \cite{BarGill2007,Yang2008} and in CARL
\cite{ColaVolpe2009}. In particular, a sub-recoil cavity linewidth
combined with a bi-chromatic pump, with frequency separated by
twice the recoil frequency, allows to observe subradiance in a
degenerate cascade between three collective momentum states
\cite{ColaBigerni2009}.

In this paper I consider a BEC in a high-finesse ring cavity
scattering photons from a pump beam into two counter-propagating
cavity modes \cite{BuxPRL2011}. Varying the pump intensity and
frequency, it is possible to populate in a controlled way a 2D
momentum lattice, where atoms belonging to different sites get
entangled with the scattered cavity photons. The paper is
organized as follow: In sec.II I derive the semiclassical 2D CARL
model for a pump beam incident at variable angle; In sec.III the
atomic motion is quantized. Sec. IV  discusses the different
quantum regimes and present some numerical result in the nonlinear
regime.

\section{Semiclassical model}
\label{sec:1}

\begin{figure}
\begin{center}
\resizebox{0.5\columnwidth}{!}{\includegraphics{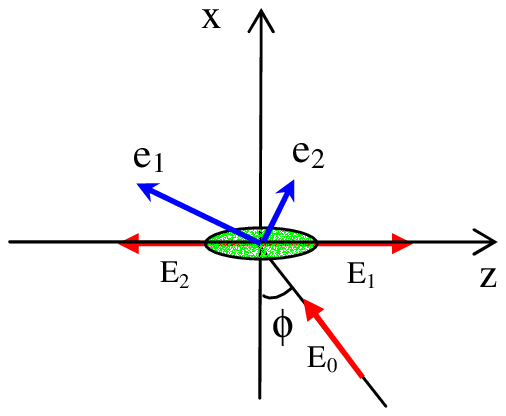}
} \caption{2D CARL configuration: the pump field $E_0$ is incident
with an angle $\phi$ with respect to the normal of the cavity axis
$\hat z$, and the cavity modes $E_1$ and $E_2$ are
counter-propagating along $\hat z$. All the fields are linearly
polarized perpendicularly to the cavity plane. The atoms recoil
along the  $\hat \mathbf{e}_1$ and $\hat \mathbf{e}_2$ directions,
when they scatter $E_0$ pump photons into the modes $E_1$ or
$E_2$, respectively.} \label{fig:1}
\end{center}
\end{figure}

Let's consider $N$ two-level atoms (with $|a\rangle$ and
$|b\rangle$ upper and lower states, respectively) in a cloud with
length $L$ and diameter $W\ll L$, exposed to an uniform,
$s$-polarized along $\hat \mathbf{e}_y$ laser beam, incident in
the cavity plane $(x,z)$ and making an angle $\phi$ respect to the
normal of the cavity's optical axis $z$ (see fig.\ref{fig:1}),
with electric field
\begin{equation}\label{Ei}
   \mathbf{ E}_i=\frac{\hat \mathbf{e}_y}{2}
   \left\{E_0e^{i(k_0x\cos\phi-k_0z\sin\phi-\omega_0 t)}+\textrm{c.c.}
   \right\}
\end{equation}
with $\omega_0=ck_0$. The pump photons are scattered in two
counter-propagating cavity modes with frequency coinciding with a
cavity eigenfrequency, $\omega_{c}=ck_{c}$, and electric field
\begin{equation}\label{Es}
   \mathbf{ E}_s=\frac{\hat \mathbf{e}_y}{2}\left\{
   E_1e^{i(k_c z -\omega_c t)}+E_2e^{-i(k_c z +\omega_c t)}+\textrm{c.c.}
   \right\}
\end{equation}

The pump and the two cavity mode fields induce the following
coherence between the states $|a\rangle$ and $|b\rangle$,
\begin{equation}\label{rho12}
   \rho_{ab}=\frac{1}{2}\left\{
   S_0e^{ik_0(x\cos\phi -z\sin\phi -ct)}+S_1e^{ik_c(z -ct)}+S_2e^{-ik_c(z +ct)}+\textrm{c.c.}
   \right\}
\end{equation}
and a force $\mathbf{F}=d_y
\mathbf{\nabla}(\mathbf{E}_i+\mathbf{E}_s)_{y}$ in the cavity
plane $(x,z)$, where $d_y=d(\rho_{ab}+\textrm{c.c.})$ is the
$y$-component of the electric dipole moment and $d$ is the dipole
matrix element. Assuming the pump-atom detuning
$\Delta_a=\omega_0-\omega_a$ much larger than the spontaneous
decay rate $\Gamma$, it is possible to see that
$S_i\approx-\Omega_i/\Delta_a$ (for $i=0,1,2$), where $\Omega_i=d
E_i/\hbar$ \cite{CARL}. A straightforward calculation shows that
the equations for the momentum components $p_{x,z}=mv_{x,z}$ are
\cite{PiovellaLP2003}
\begin{eqnarray}
  \frac{dp_x}{dt} &=& i\frac{\hbar k_0\Omega_0}{4\Delta_a}\cos\phi\left\{
  \tilde\Omega_1 e^{-i\mathbf{q}_1\cdot \mathbf{x}}+\tilde\Omega_2 e^{-i\mathbf{q}_2\cdot \mathbf{x}} -\textrm{c.c.}\right\}\label{px}\\
  \frac{dp_z}{dt} &=& -i\frac{\hbar k_0\Omega_0}{4\Delta_a}\sin\phi\left\{
  \tilde\Omega_1 e^{-i\mathbf{q}_1\cdot \mathbf{x}}-\tilde\Omega_2 e^{-i\mathbf{q}_2\cdot \mathbf{x}}
  -\textrm{c.c.}\right\}
  - i\frac{\hbar k_0\Omega_0}{4\Delta_a}\left\{
  \tilde\Omega_1 e^{-i\mathbf{q}_1\cdot \mathbf{x}}-\tilde\Omega_2 e^{-i\mathbf{q}_2\cdot \mathbf{x}}
  -\textrm{c.c.}\right\}\nonumber\\
  &-& i\frac{\hbar k_0}{2\Delta_a}\left\{
  \tilde\Omega_1\tilde\Omega_2^* e^{2ik_0z} -\textrm{c.c.}\right\}\label{pz}
\end{eqnarray}
where we assumed $k_c \sim k_0$, we introduced
$\tilde\Omega_{1,2}=\Omega_{1,2}\exp(i\Delta_c t)$, where
$\Delta_c=\omega_0-\omega_c$ is the pump-cavity detuning, and
\begin{equation}\label{q12}
    \mathbf{q}_{1,2}=k_0[\cos\phi\,\hat{\mathbf{e}}_x-(\sin\phi\pm
1)\hat{\mathbf{e}}_z],
\end{equation}
where $\mathbf{q}_{1}\cdot\mathbf{q}_{2}=0$. The equations for the
cavity mode amplitudes are
\begin{eqnarray}
  \frac{d\tilde\Omega_1}{dt} &=& \frac{ck_0d^2 n_a}{2i\epsilon_0\hbar\Delta_a}
  \left\{
   \Omega_0 \langle e^{i\mathbf{q}_1\cdot\mathbf{x}}\rangle+\tilde\Omega_1+\tilde\Omega_2 \langle e^{-2ik_0
   z}\rangle\right\}-\kappa_c\tilde\Omega_1+i\Delta_c\tilde\Omega_1\label{W1}\\
  \frac{d\tilde\Omega_2}{dt} &=& \frac{ck_0d^2 n_a}{2i\epsilon_0\hbar\Delta_a}
  \left\{
   \Omega_0 \langle e^{i\mathbf{q}_2\cdot\mathbf{x}}\rangle+\tilde\Omega_2+\tilde\Omega_1
\langle e^{2ik_0
   z}\rangle\right\}-\kappa_c\tilde\Omega_2+i\Delta_c\tilde\Omega_2\label{W2}
\end{eqnarray}
where $n_a$ is the atomic density and $\kappa_c=c{\cal{T}}/L_c$ is
the linewidth of the ring cavity with length $L_c$ and
transmission $\cal{T}$. It is more convenient to describe the
atomic motion along the directions of
$\mathbf{q}_{1,2}=q_{1,2}\hat{\mathbf{e}}_{1,2}$ with unitary
vectors $\hat{\mathbf{e}}_{1,2}$ and
$q_{1,2}=k_0\sqrt{2(1\pm\sin\phi)}$. Then, the momentum components
along these directions are, in units of the photon recoil momentum
$\hbar q_{1,2}$,
\begin{equation}\label{p12}
    p_{1,2}=\frac{\mathbf{p}\cdot\hat{\mathbf{e}}_{1,2}}{\hbar
    q_{1,2}}=\frac{k_0}{\hbar
q_{1,2}^2}\left[p_x\cos\phi- p_z(\sin\phi\pm 1)\right].
\end{equation}
Defining the phases
\begin{equation}\label{theta12}
    \theta_{1,2}=\mathbf{q}_{1,2}\cdot
    \mathbf{x}=k_0x\cos\phi-
k_0z(\sin\phi\pm 1)
\end{equation}
and the dimensionless  field amplitudes
\begin{equation}\label{a12}
    a_{1,2}=i\sqrt{\frac{\epsilon_0\hbar
    V}{2 d^2\omega_0}}\tilde\Omega_{1,2}=i\sqrt{\frac{\epsilon_0
    V}{2\hbar\omega_0}}E_{1,2}e^{i\Delta_c t}
\end{equation}
where $V$ is the interaction volume, the complete equations for
$N$ atoms and the two cavity mode amplitudes are
\begin{eqnarray}
\frac{d\theta_{1j}}{dt} &=& 2\omega_{r1}p_{1j}\label{eq1}\\
\frac{d\theta_{2j}}{dt} &=& 2\omega_{r2}p_{2j}\label{eq2}\\
\frac{dp_{1j}}{dt} &=& \frac{g\Omega_0}{2\Delta_a}\left\{
  a_1 e^{-i\theta_{1j}}+\textrm{c.c.}\right\}+i\frac{g^2}{\Delta_a}\left\{
  a_1 a_2^* e^{i(\theta_{2j}-\theta_{1j})}-\textrm{c.c}.\right\}\label{eq3}\\
  \frac{dp_{2j}}{dt} &=& \frac{g\Omega_0}{2\Delta_a}\left\{
  a_2 e^{-i\theta_{2j}}+\textrm{c.c.}\right\}-i\frac{g^2}{\Delta_a}\left\{
  a_1 a_2^* e^{i(\theta_{2j}-\theta_{1j})}-\textrm{c.c.}\right\}\label{eq4}\\
   \frac{da_1}{dt} &=& \frac{Ng\Omega_0}{2\Delta_a}\langle
   e^{i\theta_1}\rangle
   -i\frac{Ng^2}{2\Delta_a}a_2\langle
e^{-i(\theta_2-\theta_1)}\rangle-\kappa_c a_1+i\left(\Delta_c-\frac{Ng^2}{\Delta_a}\right)a_1\label{eq5}\\
\frac{da_2}{dt} &=& \frac{Ng\Omega_0}{2\Delta_a}\langle
   e^{i\theta_2}\rangle
   -i\frac{Ng^2}{2\Delta_a}a_1\langle
e^{i(\theta_2-\theta_1)}\rangle-\kappa_c
a_2+i\left(\Delta_c-\frac{Ng^2}{\Delta_a}\right)a_2\label{eq6}
\end{eqnarray}
where $j=1,\dots,N$, $g=\sqrt{d^2\omega_0/(2\epsilon_0\hbar V)}$
is the single-photon Rabi frequency and
\begin{equation}\label{omegarec}
    \omega_{r1,2}=\frac{1}{2}\left(1\pm\sin\phi\right)\omega_{r},
\end{equation}
where $\omega_{r}=2\hbar k_0^2/m$ is the maximum photon recoil
frequency. Notice that the atoms move along the recoiling
directions $\hat \mathbf{e}_1$ or $\hat \mathbf{e}_2$ (see
fig.\ref{fig:1}) when they scatter the pump photons into the
cavity modes $a_1$ and $a_2$, respectively (see the first terms on
the right hand sides of Eqs.(\ref{eq3}) and (\ref{eq4}),
representing the dipole forces depending on $\theta_{1j}$ and
$\theta_{2j}$, respectively). Furthermore, the atoms recoil
further along the cavity axis $\hat z$ when they exchange photons
between the two cavity modes themselves (see the second terms on
the right hand sides of Eqs.(\ref{eq3}) and (\ref{eq4}),
representing the dipole force due to the two-cavity mode
interference and depending on $\theta_{2j}-\theta_{1j}=2k_0z_j$).
Notice that the longitudinal dipole force breaks the pump-atom
detuning $\Delta_a$ symmetry: In fact, if $\Delta_a\rightarrow
-\Delta_a$ and $a_{1,2}\rightarrow -a_{1,2}$, these terms change
sign too (together with the collective single-photon light shift
$Ng^2/\Delta_a$).

\section{Quantum model}
\label{sec:2}

In a quantum theory, the classical variables $\theta_{1j}$,
$\theta_{2j}$, $p_{1j}$, $p_{2j}$, $a_1$ and $a_2$ are promote to
operators, with commutation rules $[\theta_{\alpha j},p_{\beta
m}]=i\delta_{\alpha\beta}\delta_{jm}$ and
$[a_\alpha,a_\beta^\dagger]=\delta_{\alpha\beta}$ where
$\alpha,\beta=1,2$. Without cavity losses (i.e. $\kappa_c=0$),
Eqs.(\ref{eq1})-(\ref{eq6}) derive by the following Hamiltonian:
\begin{eqnarray}
  H &=& \sum_{j=1}^N\left\{
  \omega_{r1}p_{1j}^2+\omega_{r2}p_{2j}^2+i\frac{g\Omega_0}{2\Delta_a}
  \left[a_1^\dagger e^{i\theta_{1j}}+a_2^\dagger
  e^{i\theta_{2j}}-\textrm{h.c.}\right]+
  \frac{g^2}{\Delta_a}
  \left[a_1a_2^\dagger e^{i(\theta_{2j}-\theta_{1j})}+\textrm{h.c.}
  \right]\right\}\nonumber\\
  &-&\left(\Delta_c-\frac{Ng^2}{\Delta_a}\right)(a_1^\dagger a_1+a_2^\dagger
  a_2).\label{ham}
\end{eqnarray}
The single-particle Hamiltonian
$H_1(\theta_1,\theta_2,p_1,p_2,a_1,a_1^\dagger,a_2,a_2^\dagger)$
can be second-quantized as
\begin{eqnarray}
  \hat H &=& \int_0^{2\pi}d\theta_1\int_0^{2\pi}d\theta_2\hat\Psi^\dagger(\theta_1,\theta_2)
  H_1(\theta_1,\theta_2,-i\partial_{\theta_1},-i\partial_{\theta_2},a_1,a_1^{\dagger},a_2,a_2^{\dagger})
  \,\hat\Psi(\theta_1,\theta_2)\quad\quad\label{ham2}
\end{eqnarray}
where  the quantum field operator $\hat\Psi(\theta_1,\theta_2)$
obeys bosonic equal-time commutation rules
$[\hat\Psi(\theta_1,\theta_2),\hat\Psi^\dagger(\theta_1',\theta_2')]=\delta(\theta_1-\theta_1')\delta(\theta_2-\theta_2')$
and
$[\hat\Psi(\theta_1,\theta_2),\hat\Psi(\theta_1',\theta_2')]=0$,
with normalization condition $\int d\theta_1\int
d\theta_2\hat\Psi^\dagger(\theta_1,\theta_2)\hat\Psi(\theta_1,\theta_2)=\hat
N$. Introducing the annihilation operators for the two momentum
components $p_1$ and $p_2$, i.e.
$\hat\Psi(\theta_1,\theta_2)=\sum_{m,n}\hat
c_{m,n}u_m(\theta_1)u_n(\theta_2)$, where
$u_m(\theta_{1,2})=(1/\sqrt{2\pi})\exp[im\theta_{1,2}]$ and $[\hat
c_{m,n},\hat c_{m',n'}^\dagger]=\delta_{m,m'}\delta_{n,n'}$, the
Heisenberg equations for $\hat c_{m,n}$, $\hat a_1$ and $\hat a_2$
read:
\begin{eqnarray}
  \frac{d\hat c_{m,n}}{dt} &=& -i\left[m^2\omega_{r1}+ n^2\omega_{r2}\right]\hat c_{m,n}\nonumber\\
  &+& \frac{g\Omega_0}{2\Delta_a}
  \left[\hat a_1^\dagger\hat c_{m-1,n}+\hat a_2^\dagger\hat c_{m,n-1}-\hat a_1\hat c_{m+1,n}-\hat a_2\hat c_{m,n+1}\right]\nonumber\\
  &-&i\frac{g^2}{\Delta_a}\left[a_1a_2^\dagger\hat c_{m+1,n-1}+a_1^\dagger a_2\hat c_{m-1,n+1}\right]\label{cmn:before}\\
  \frac{d\hat a_1}{dt} &=&  \frac{g\Omega_0}{2\Delta_a}\sum_{m,n}\hat c^\dagger_{m,n}\hat c_{m-1,n}
  -i\frac{g^2}{\Delta_a}\hat a_2\sum_{m,n}\hat c_{m,n}^\dagger\hat c_{m-1,n+1}+i\Delta\hat a_1\label{a1}\\
  \frac{d\hat a_2}{dt} &=&  \frac{g\Omega_0}{2\Delta_a}\sum_{m,n}\hat c^\dagger_{m,n}\hat
  c_{m,n-1}
  -i\frac{g^2}{\Delta_a}\hat a_1\sum_{m,n}\hat  c_{m,n}^\dagger\hat c_{m+1,n-1}+i\Delta
  a_2\label{a2}
\end{eqnarray}
where $\Delta=\Delta_c-N g^2/\Delta_a$. Notice that in
Eq.(\ref{cmn:before}) we have neglected the global phase factor
proportional to $\Delta(a_1^\dagger a_1+a_2^\dagger a_2)$.

In the following we will neglect the quantum nature of the
operators $\hat c_{m,n}$ and $\hat a_{1,2}$ and we treat them as
complex dynamical variables. Furthermore, we introduce
dimensionless time, $\tau=2\omega_{r}\rho t$, field amplitudes,
$A_{1,2}=(a_{1,2}/\sqrt{\rho N})\exp(-i\Delta t)$, and pump
parameter, $A_{0}=a_{0}/\sqrt{\rho N}$, where $a_0=\Omega_0/(2g)$
is such that $a_0^2$ is the pump photon number. The collective
CARL parameter $\rho$ is defined as \cite{CARL}
\begin{equation}\label{rho}
    \rho=\left(\frac{a_0 g^2\sqrt{N}}{2\Delta_a\omega_{r}}\right)^{2/3},
\end{equation}
Then, defining $C_{m,n}=(1/\sqrt{N})c_{m,n}\exp[i(m+n)\Delta t]$,
$\kappa=\kappa_c/(2\omega_{r}\rho)$ and
$\delta=\Delta/\omega_{r}$, Eqs.(\ref{cmn:before})-(\ref{a2})
yield
\begin{eqnarray}
  \frac{dC_{m,n}}{d\tau} &=& -\frac{i}{2\rho}\left[m^2\left(\frac{1+\sin\phi}{2}\right)+n^2\left(\frac{1-\sin\phi}{2}\right)
  -(m+n)\delta\right]C_{m,n}\nonumber\label{eq:cmn}\\
  &+&\rho
  \left[A_1^*C_{m-1,n}+A_2^*C_{m,n-1}-A_1C_{m+1,n}-A_2C_{m,n+1}\right]\nonumber\\
  &-&i\frac{\rho}{A_0}\left[A_1A_2^*C_{m+1,n-1}+A_1^* A_2C_{m-1,n+1}\right]\\
  \frac{dA_1}{d\tau} &=&  \sum_{m,n}C^*_{m,n}C_{m-1,n}
  -i\frac{A_2}{A_0}\sum_{m,n}C_{m,n}^*C_{m-1,n+1}-\kappa A_1\label{eq:A1}\\
  \frac{dA_2}{d\tau} &=&  \sum_{m,n}C^*_{m,n}C_{m,n-1}
  -i\frac{A_1}{A_0}\sum_{m,n}C_{m,n}^*C_{m+1,n-1}-\kappa
  A_2\label{eq:A2}
\end{eqnarray}
Notice that the total probability is conserved, i.e.
$\sum_{m,n}|C_{m,n}|^2=1$. The growth rates for the two modes are
$G_{1,2}=-2\textrm{Im}(\lambda)(2\omega_{r}\rho)$, where $\lambda$
is the solution of the cubic dispersion relation:
\begin{equation}\label{cubic}
    \left(\lambda-\frac{\delta}{2\rho}-i\kappa\right)\left[\lambda^2-\left(\frac{s_{1,2}}{2\rho}\right)^2\right]+s_{1,2}=0,
\end{equation}
where $s_{1,2}=(1\pm\sin\phi)/2$. Notice that the longitudinal
lattice term is nonlinear (being proportional to $A_1A_2^*$) and
it does not contribute to the dispersion relation (\ref{cubic}).
In the following we will indicate the 2D momentum lattice states
as $(m,n)$, associated with momentum components
$\mathbf{p}_1=m(\hbar q_1)\hat \mathbf{e}_1$ and
$\mathbf{p}_2=m(\hbar q_2)\hat \mathbf{e}_2$, respectively. In the
linear regime, the two cavity modes grow independently and atoms
(initially in $(0,0)$) populate the states $(\pm 1,0)$ and $(0,\pm
1)$, respectively. In particular, when the laser beam is parallel
to the cavity axis $\hat z$ (i.e. $\phi=90^\circ$), $s_1=1$ and
$s_2=0$: Only the mode $a_1$ grows, so we can set $a_2=0$ and the
model reduces to the usual 1D CARL, with $c_{m,n}=\delta_{n,0}c_m$
\cite{Gatelli2001}.

\section{Discussion}
\label{sec:3}
After scattered a photon with momentum $\hbar
\mathbf{k}$ and energy $\hbar\omega$, the atom recoils with
momentum $\mathbf{p}$ determined by energy and momentum
conservation laws, i.e. $\hbar
\mathbf{k}_0=\hbar\mathbf{k}+\mathbf{p}$ and
$\hbar\omega_0=\hbar\omega+p^2/2m$, so that $p_{x}=\hbar
k_0\cos\phi$, $p_{z}=\hbar k_0(\sin\phi\mp 1)$ and
$\omega=\omega_0-\omega_{r}(1\pm\sin\phi)/2$, where the upper and
lower sign is for a photon emitted in the cavity mode $a_1$ or
$a_2$, respectively. Since $\omega$ is near the cavity mode
frequency $\omega_c$, then by tuning the pump frequency $\omega_0$
near the cavity mode frequency $\omega_c$ it is possible to
enhance one mode with respect to the other, depending on incidence
angle $\phi$, gain bandwidth $\omega_r\rho$ and cavity linewidth
$\kappa_c$ values \cite{BuxPRL2011}.

\subsection{'Good-Cavity' and 'Superradiant' regimes}
\label{sec:3:1} The cubic dispersion relation (\ref{cubic})
provides the expression for the gain rates $G_{1,2}$ of the two
cavity modes in the good-cavity (GC) regime (i.e. when
$2\kappa_c\ll G_{1,2}$) and in the super-radiant (SR) regime (i.e.
when $2\kappa_c\ge G_{1,2}$), either in the semiclassical regime
(i.e. when $\Delta\omega\gg\omega_r$) or in the quantum regime
(i.e. when $\Delta\omega\le\omega_r$), where $\Delta\omega$ is the
resonant gain bandwidth
\cite{Gatelli2001,Martinucci2002,Piovella2003}. In particular, in
the quantum regime the atoms scatter the pump scattered photons
only forward, since the atomic recoil red-shifts the scattered
photon frequency (i.e. at $\omega-\omega_{r1,2}$) such that it is
set out of the resonant gain bandwidth \cite{Gatelli2001}. So, in
the quantum regime the atoms populate initially only the
positive-momentum states $(1,0)$ and $(0,1)$. In the GC limit we
can set $\kappa_c=0$, obtaining
$G_{1,2}=2\sqrt{(2\rho)^{3}\omega_r^2-\Delta_{1,2}^2}$ for
$\Delta_{1,2}< (2\rho)^{3/2}\omega_r$, where
$\Delta_{1,2}=\Delta-\omega_{r1,2}$ is the detuning taking into
account the recoil shift. Hence, in the quantum GC regime, the
maximum gain and the gain bandwidth are
$G_{max}=2(2\rho)^{3/2}\omega_r$ and
$\Delta\omega=2(2\rho)^{3/2}\omega_r=G_{max}$, respectively, and
the conditions necessary to observe it are $\rho<1$ and
$\kappa_c\ll\omega_r$. In the quantum SR regime,
$G_{1,2}=\kappa_c(2\omega_r\rho)^2/(\Delta_{1,2}^2+\kappa_c^2)$,
so that the maximum gain is $G_{max}=(2\omega_r\rho)^2/\kappa_c$
and the gain bandwidth is $\kappa_c$ \cite{Martinucci2002}. The
condition necessary to observe the quantum SR regime is
$\omega_r\rho\le\kappa_c<\omega_r$. Notice that in the quantum
regime gain and bandwidth are the same for the two modes. However,
increasing $\rho$ for a given cavity linewidth $\kappa_c$, the
system moves toward the classical GC limit, $G_{1,2}\gg\omega_r$,
where the recoil shift can be neglected and the gain is centered
around $\Delta=0$, with
$G_{max}=2\sqrt{3}\,\omega_r\rho(s_{1,2})^{1/3}$. Hence, in the
classical regime the two cavity modes have different gain rates.
As an example of an intermediate case (with parameters close to
those of ref.\cite{BuxPRL2011}), fig.\ref{fig:2} shows the gain
$G_{1,2}$ (in CARL bandwidth $2\rho\omega_r$ units) vs. the pump
cavity detuning $\delta=\Delta/\omega_r$, for $\phi=45^\circ$,
$\rho=0.4$ and $\kappa_c=0.8\omega_r$.
\begin{figure}
\begin{center}
\caption{Gain coefficients $g_{1,2}=-2Im(\lambda)$ (continuous
blue line for mode $1$ and dashed red line for mode $2$) vs. the
pump-cavity detuning $\delta=\Delta/\omega_r$, for
$\phi=45^\circ$, $\rho=0.4$ and $\kappa_c=0.8\omega_r$.}
\label{fig:2}
\end{center}
\end{figure}

\subsection{Symmetric nonlinear regime}
\label{sec:3:2} The case where the laser beam is perpendicular to
the cavity axis $\hat z$ (i.e. $\phi=0$ and $s_{1,2}=1/2$) has
been discussed in details in ref.\cite{PiovellaLP2003}. Here, the
two modes are symmetric and atoms move at $45^\circ$ forward and
backward with respect to the cavity axis. This is also the
original configuration of the Superradiant Rayleigh scattering
experiment of ref.\cite{Inouye1999}. Following
ref.\cite{PiovellaLP2003}, it results that in the quantum regime
atoms populate sequentially the momentum states $(n,n)$ (with
$n=1,2,\dots$) by a four-level 'diamond' transition, passing
through the intermediate states $(n-1,n)$ and $(n,n-1)$. Since
$c_{n,m}=c_{m,n}$ and $A_1=A_2$, each transition
$(n-1,n-1)\rightarrow (n,n)$ can be described by optical Bloch
equations for two-level systems once a population difference
$W=|c_{n-1,n-1}|^2-|c_{n,n}|^2$ and a polarization
$S=c_{n-1,n-1}^*c_{n-1,n}+c_{n-1,n}^*c_{n,n}$ are introduced. In
the quantum SR regime and at resonance (i.e. for
$\Delta=\omega_r/2$), the populations evolve as
$|c_{n,n}(\tau)|^2=(1/4)\left\{1-\tanh[\sqrt{\rho/\kappa}(\tau-\tau_D)]\right\}^2$
and
$|c_{n-1,n}(\tau)|^2=(1/4)\textrm{sech}^2[\sqrt{\rho/\kappa}(\tau-\tau_D)]$,
whereas the cavity photon number  is $\langle
n_{ph}\rangle=N\textrm{sech}^2[\sqrt{\rho/\kappa}(\tau-\tau_D)] $,
where $\tau_D\approx\sqrt{\rho}\ln(\rho)$. Notice that the maximum
occupation probability of the intermediate states $(n-1,n)$ and
$(n,n-1)$ is $1/4$. This configuration is particular attractive,
since either entanglement \cite{Porras2008} and subradiance could
be there easily addressed, as discussed by Crubellier et al.
\cite{Crubellier1985,Crubellier1986}.

\subsection{Asymmetric nonlinear regime}
\label{sec:3:3} 
\begin{figure}
\begin{center}
\resizebox{0.85\columnwidth}{!}{\includegraphics{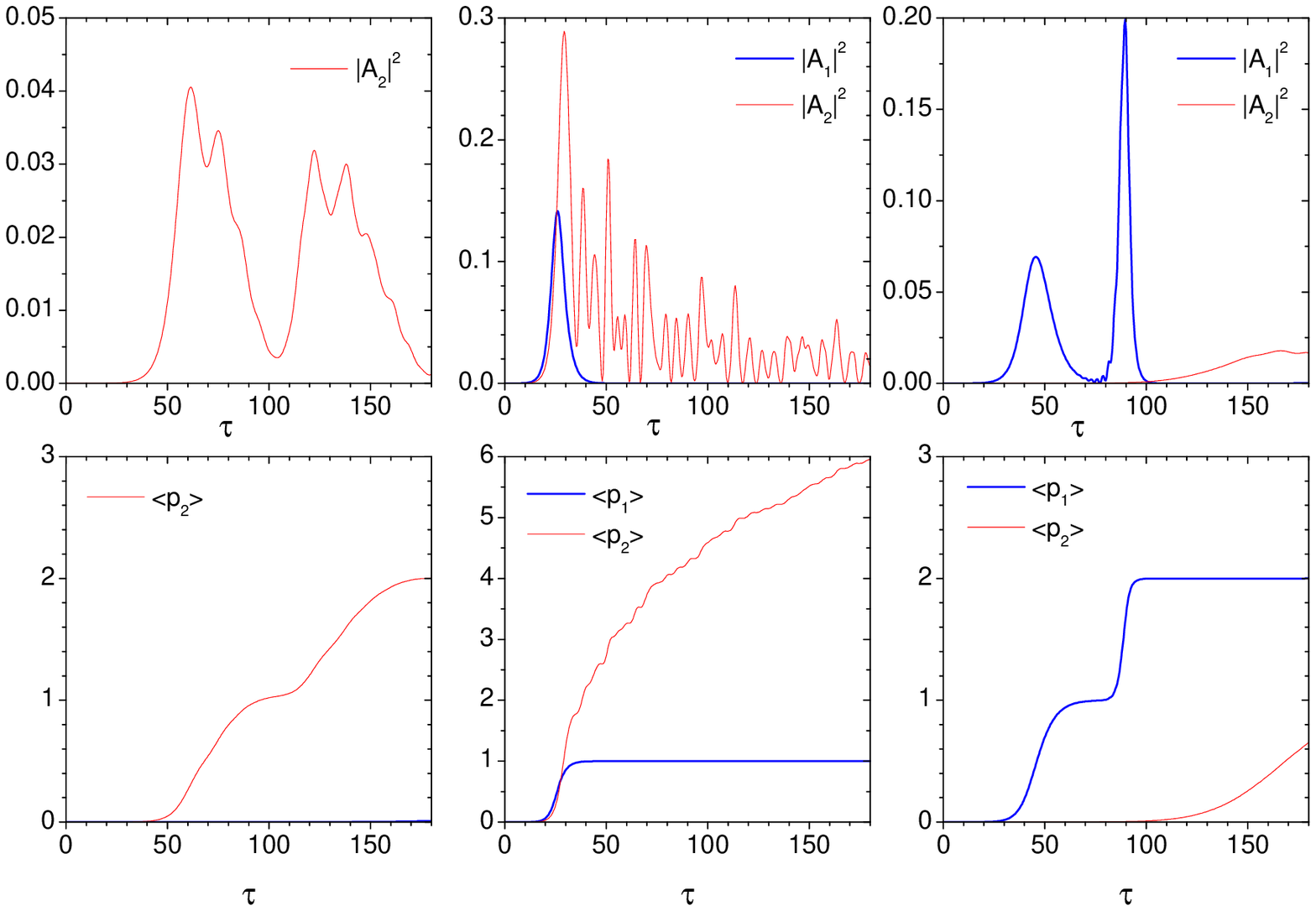}}
\caption{$|A_{1,2}|^2$ (upper line) and $\langle p_{1,2}\rangle$
(lower line) vs. dimensionless time  $\tau$ for
$\Delta=-2.5\omega_r$ (left column), $\Delta=-0.025\omega_r$
(central column) and $\Delta=2\omega_r$ (right column), for
$\phi=45^\circ$, $\rho=0.4$ and $\kappa_c=0.8\omega_r$. Blue thick
lines are for the mode $1$, red thin lines for the mode $2$.}
\label{fig:3}
\end{center}
\end{figure}

The symmetry of the case with perpendicular
incidence is broken when the laser beam shines the atoms with an
oblique incidence angle, as can be seen for instance in
fig.\ref{fig:1}. In this case, changing the laser frequency
$\omega_0$ with respect to the cavity mode frequency $\omega_c$
allows to unbalance the two counter-propagating cavity modes, as
well as the two momentum components $p_1$ and $p_2$. Furthermore,
nonlinearity induces more complicated dynamical structures
resulting from the interplay of cooperative gain and cavity
losses, when more than one photon is scattered by the condensate.
As an example, we consider the case with $\phi=45^\circ$,
$\rho=0.4$, $\kappa=0.8\omega_r$ and different pump-cavity
detuning $\Delta$. In order to ge the analysis simpler, we neglect
the longitudinal lattice (i.e. the last term in the right-hand
side of Eqs.(\ref{eq:cmn}) and the second terms in the right-hand
side of Eqs.(\ref{eq:A1}) and (\ref{eq:A2})) assuming $A_{1,2}\ll
A_0$. Fig.\ref{fig:3} shows the result of numerical integration of
Eqs.(\ref{eq:cmn})-(\ref{eq:A2}) for different pump-cavity
detuning values: $\Delta=-2.5\omega_r$ (left column),
$\Delta=-0.025\omega_r$ (central column) and $\Delta=2\omega_r$
(right column); mode intensities $|A_{1,2}|^2$
 and the average momentum components $\langle
p_{1,2}\rangle$ are shown vs. $\tau$ in the upper and lower lines,
respectively; blue thick lines refer to mode $1$ and red thin
lines refer to mode $2$.
For $\Delta=-2.5\omega_r$ (left column) $G_1\ll G_2$, so that the
mode $1$ does not grow appreciably. The atoms move along the
$\hat{\mathbf{e}}_2$ direction, up to the momentum state $(0,2)$
after a time $\tau=180$. For $\Delta=-0.025\omega_r$ (central
column) the gain rates are equal ($G_1=G_2=0.41\omega_r$) and the
atoms initially equally populate the momentum states $(1,0)$ and
$(0,1)$. However, later on the atoms turn toward the
$\hat{\mathbf{e}}_2$ direction, populating the states $(1,m)$ with
$m=1,\dots,6$ after a time $\tau=180$. This rather peculiar
behavior has been observed also in the experiment of
ref.\cite{BuxPRL2011}. This behavior can be easily understood
observing that the recoil shift for the mode $1$ is $s_1/s_2=5.83$
times larger than for the mode $2$, so that the incident photons
are set out of resonance after the atoms have scattered the first
laser photon into the mode $1$, whereas the incident photons
remain well inside the resonant gain bandwidth when scattered into
the mode $2$ (see fig.\ref{fig:2}). As a consequence, the atoms
scatter a single photon into the mode $1$, stopping at the
momentum states $(m,n)$ with $m=1$, whereas they are allowed to
scatter photons into the mode $2$ (up to $n=6$, as shown in
central column of fig.\ref{fig:3}), populating the momentum states
$(1,n)$. Finally, for $\Delta=2\omega_r$ (right column of
fig.\ref{fig:3})) $G_2\ll G_1$, so initially only the mode $1$
grows and atoms populate sequentially the states $(1,0)$ and
$(2,0)$; however, at a longer time the mode $2$ grows and reaches
saturation, so that the atoms populate also the state $(2,1)$.

\section{Conclusions}
\label{sec:4} I have derived the semiclassical and quantum model
of the collective atomic recoil laser (CARL) for a Bose-Einstein
condensate set in an arm of an high-finesse ring cavity, with a
laser beam incident at an oblique angle with respect to the cavity
axis. The atoms scatter photons into two counterpropagating cavity
modes, recoiling along two different directions determined by the
incidence angle. For perpendicular incidence, atoms scatter
symmetrically the pump photons into the two cavity modes,
populating sequentially symmetric momentum states $(n,n)$ with
$n=1,2,\dots$. Conversely, for oblique incidence it is possible to
populate in a controlled way different momentum states by tuning
the pump frequency, as experimentally done in
ref.\cite{BuxPRL2011}. Similarly to the 1D geometry
\cite{Piovella2003,Cola2004}, it is expected that atoms belonging
to different momentum states may be entangled between themselves
and/or with the photons scattered in the cavity modes.
Furthermore, an even reacher scenario can be realized by a
bichromatic pump with frequency spacing tuned around the recoil
frequency, which is expected to enhance or inhibit the transfer of
atoms between different momentum states
\cite{ColaVolpe2009,ColaBigerni2009}.

\section{Acknowledgments}

This work is dedicated to Federico Casagrande, in memory of our
long-standing friendship and of his unforgettable kindness and
sympathy. I would like to thank Simone Bux, Philippe W. Courteille
and Claus Zimmermann for helpful discussions about the experiment
described in ref.\cite{BuxPRL2011}. This work has been supported
by the Research Executive Agency (program COSCALI No.
PIRSES-GA-2010-268717).

\end{document}